\documentclass[5p,twocolumn,times]{elsarticle}
\usepackage{lipsum}
\usepackage{lineno,hyperref}
\modulolinenumbers[5]
\usepackage[pdftex]{color}
\usepackage[font=footnotesize,labelfont=bf]{caption}
\usepackage[font=footnotesize,labelfont=bf]{subcaption}
\journal{Journal of \LaTeX\ Templates}
\usepackage{colortbl}
\usepackage{xcolor}

\usepackage{amssymb}

\biboptions{compress}

\usepackage[figuresright]{rotating}

\begin{document}

\begin{frontmatter}


\title{Aggregation as an antipredator strategy in the rock-paper-scissors model}


\address[1]{Escola de Ci\^encias e Tecnologia, Universidade Federal do Rio Grande do Norte\\
Caixa Postal 1524, 59072-970, Natal, RN, Brazil}
\address[2]{Institute for Biodiversity and Ecosystem
Dynamics, University of Amsterdam, Science Park 904, 1098 XH
Amsterdam, The Netherlands}
\address[3]{Departamento de Engenharia Biomédica, Universidade Federal do Rio Grande do Norte\\
  Av. Senador Salgado Filho, 300 Lagoa Nova,
CEP 59078-970 - Natal, RN - Brazil}
\address[4]{Edmond and Lily Safra International Neuroscience Institute, Santos Dumont Institute\\
Av Santos Dumont, 1560, 59280-000, Macaiba, RN, Brazil}

\author[1,2]{J. Menezes}  
\author[1]{E. Rangel} 
\author[3,4]{B. Moura} 

\begin{abstract}
We study a nonhierarchical tritrophic system, whose predator-prey interactions are described by the rock-paper-scissors game rules. In our stochastic simulations, individuals may move strategically towards the direction with more conspecifics to form clumps instead of moving aimlessly on the lattice.  Considering that the conditioning to move gregariously depends on the organism's physical and cognitive abilities, we introduce a maximum distance an individual can perceive the environment and a minimum conditioning level to perform the gregarious movement. We investigate the pattern formation and compute the average size of the single-species spatial domains emerging from the grouping behaviour. The results reveal that the defence tactic reduces the predation risk significantly, being more profitable if individuals perceive further distances, thus creating bigger groups. Our outcomes show that the species with more conditioned organisms dominate the cyclic spatial game, controlling most of the territory. On the other hand, the species with fewer individuals ready to perform aggregation strategy gives its predator the chance to fill the more significant fraction of the grid.
The spatial interactions assumed in our numerical experiments constitute a data set that may help biologists and data scientists understand how local interactions influence ecosystem dynamics. 
\end{abstract}


\end{frontmatter}


\section{Introduction}
\label{sec:int}

Predator-prey interactions are responsible for the stability
of the rich biodiversity found in nature\cite{ecology,Nature-bio,BUCHHOLZ2007401}. 
In evolutionary biology, investigating organisms' behaviour is central
to understanding how to manage the conservation of ecosystems\cite{doi:10.1098/rstb.2019.0012}. For example, antipredator behaviour has been observed in many species, from invertebrates to vertebrates \cite{Thanatonis,ContraAtacck2}. It has been reported that
the success of the response to an imminent predator's attack depends on the organisms' ability to detect a nearby enemy and the energy expended by each teammate in the collective action\cite{Cost2,olfactory2,LizardB1,detection}. 
For this reason, many animals live in groups; thus, reducing the chances of being consumed in an eventual predator attack \cite{manyeyes,dilution1, dilution2,Grouping1,Grouping2,fishing,strategy1,strategy4,strategy5,strategy3}.

Rock-paper-scissors game rules have successfully modelled the nonhierarchical cyclic interactions found in many biological systems \cite{Allelopathy,Coli,lizards,Extra1,Avelino-PRE-86-036112,Szolnoki_2020,uneven,Moura}. This has been allowed researchers to discover mechanisms leading to the emergence of spatial patterns which controls population dynamics in scenarios where cyclic predator-prey relationships are present \cite{weakest,doi:10.1021/ja01453a010,Volterra,PhysRevE.78.031906,0295-5075-121-4-48003,PhysRevE.82.066211,Rev6}. 
In Ref. \cite{Rev1}, the authors study the main aspects of the cyclic evolutionary games in the generalised rock-paper-scissors game in structured populations, showing that mobility plays a central role in promoting or jeopardising biodiversity. It has also been shown that physical constraints, movement strategies, and the breaking of the unidirectional invasions can unbalance the cyclic nonhierarchical game, impacting pattern formation and affecting coexistence \cite{Rev4,Moura,Breaking}. 
Furthermore, cyclic dominance plays a fundamental role in the spatial interactions in social systems, public good with punishment, and human bargaining \cite{Rev2,Rev3}.

Recently, the role of antipredator behaviour has been explored 	in spatial simulations of the rock-paper-scissors model, 
revealing the emergence of spatial patterns \cite{Anti1,anti2}. It has been demonstrated that the reduction in the predation risk is accentuated if the antipredator reaction is less localised, demanding less energy from each organism participating in the collective strategy.  This work investigates the aggregation behaviour as an antipredator strategy in nonhierarchical tritrophic systems described by the spatial rock-paper-scissors game rules. Performing spatial stochastic simulations, we consider that organisms can scan the environment to be aware of their conspecifics. To minimise the chances of being preyed on, the organism moves gregariously towards the direction with 
the larger number of conspecifics.

In addition, we introduce a conditioning factor to implement the individual physical and cognitive ability to perform the directional self-preservation movement. We implement a maximum distance an individual can perceive its neighbourhood; thus, we study how the organism's perception radius controls pattern formation and influences the cyclic species territorial dominance. We also aim to discover how aggregation behaviour impacts the predation risk in scenarios where not all organisms are conditioned to perform the gregarious movement. To this purpose, we further explore the dynamics of densities of species in uneven scenarios where organisms of one out of the species are more or less conditioned than the other species. 


\begin{figure}[t]
\centering
\includegraphics[width=40mm]{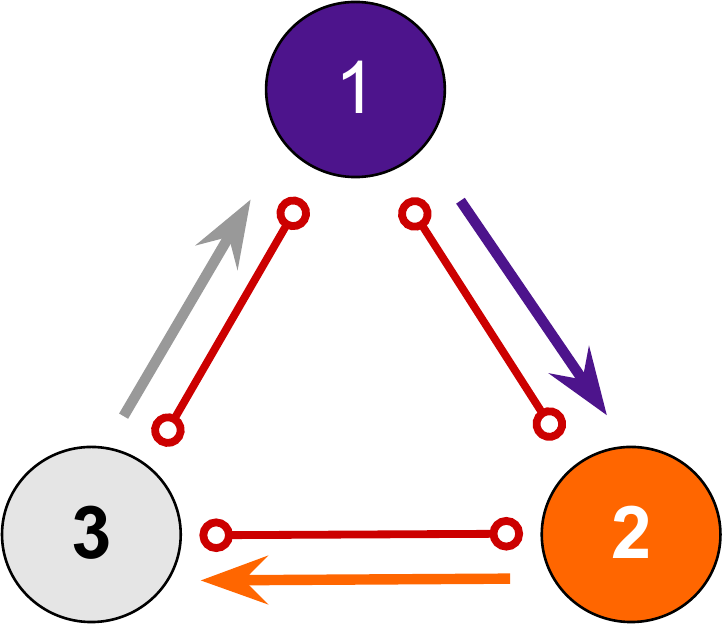}
\caption{Illustration of the predation and mobility rules in our model. The arrows indicate the throphic dominance between two species, while the
red bars show that two organisms of any species can switch positions during a mobility interaction.
}
\label{fig1}
\end{figure}
\begin{figure}[t]
\centering
\includegraphics[width=40mm]{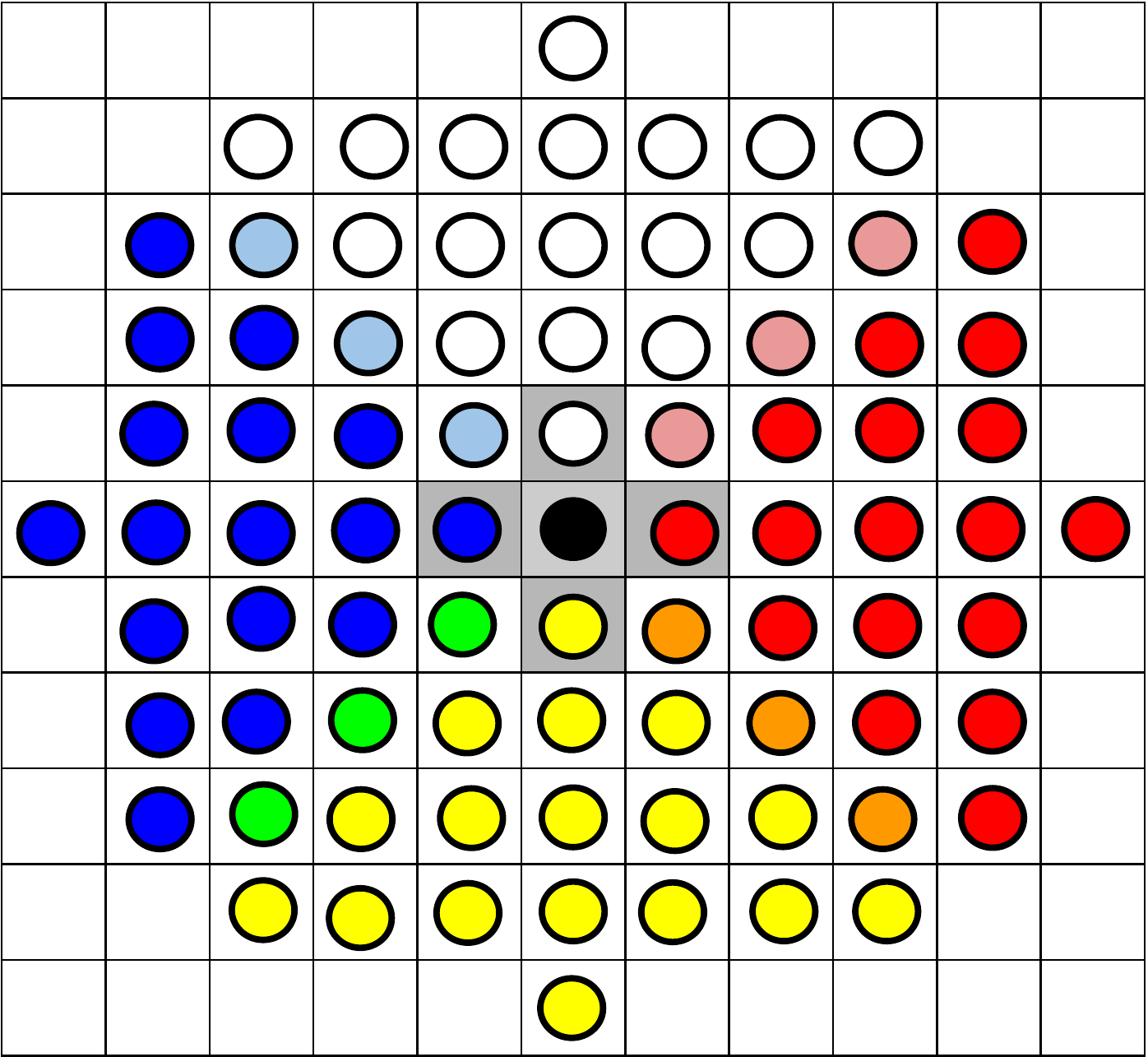}
\caption{Illustration of the numerical implementation
of the organisms' neighbourhood perception when moving directionally.
Positioned at the black grid site, an individual moves to one of the next grid points (dots with grey background) in the direction with more conspecifics. For making the decision, the organism scans its conspecifics in the grid sites on the north (white, light blue, and light red), south (yellow, green, and orange), east (red, pink, and orange), and west (blue, green, and light blue) directions. 
}
\label{fig1}
\end{figure}
\begin{figure*}
\centering
    \begin{subfigure}{.23\textwidth}
        \centering
        \includegraphics[width=40mm]{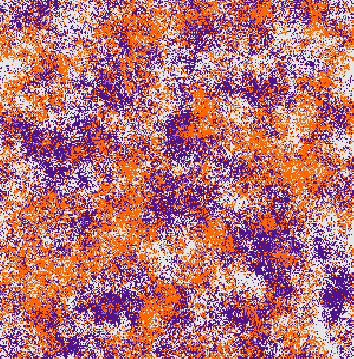}
        \caption{}\label{fig2a}
    \end{subfigure} %
       \begin{subfigure}{.23\textwidth}
        \centering
        \includegraphics[width=40mm]{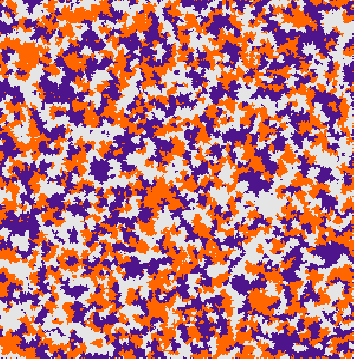}
        \caption{}\label{fig2b}
    \end{subfigure} %
   \begin{subfigure}{.23\textwidth}
        \centering
        \includegraphics[width=40mm]{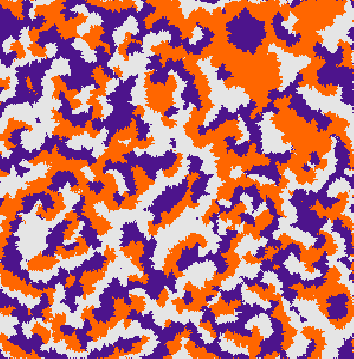}
        \caption{}\label{fig2d}
    \end{subfigure} 

\caption{Spatial patterns obtained from a lattice with $300^2$ grid points running until $3000$ generations. Figures a, b, and c show the organisms' spatial distribution for $\mathcal{R}=0$
 (standard model), $\mathcal{R}=3$, and $\mathcal{R}=7$, respectively. The colours follow the scheme in Fig. 1.}
  \label{fig2}
\end{figure*}
\begin{figure}
 \centering
    \begin{subfigure}{.4\textwidth}
        \centering
        \includegraphics[width=75mm]{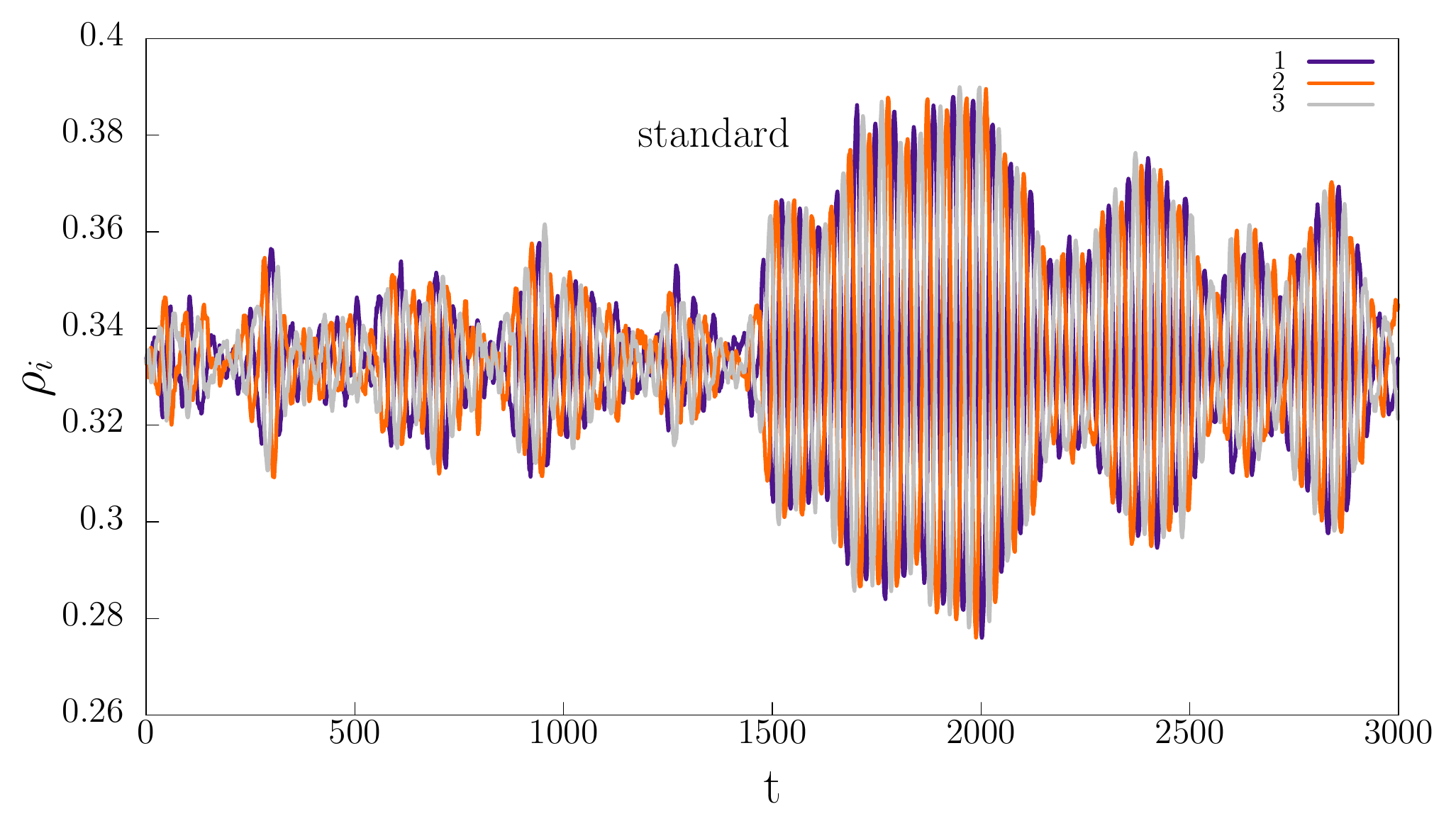}
        \caption{}\label{fig2a}
    \end{subfigure}\\
       \begin{subfigure}{.4\textwidth}
        \centering
        \includegraphics[width=75mm]{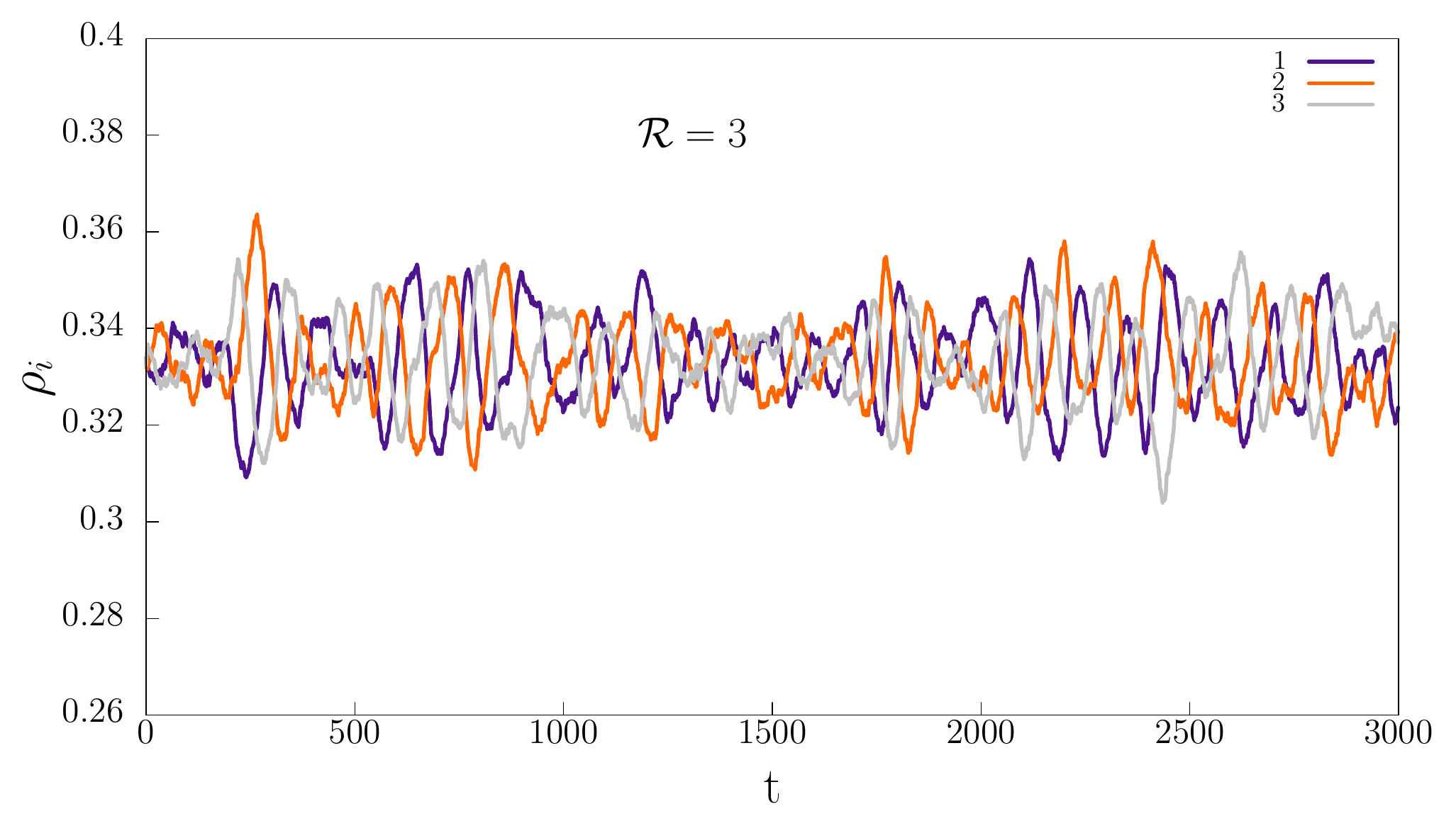}
        \caption{}\label{fig2b}
    \end{subfigure}\\
   \begin{subfigure}{.4\textwidth}
        \centering
        \includegraphics[width=75mm]{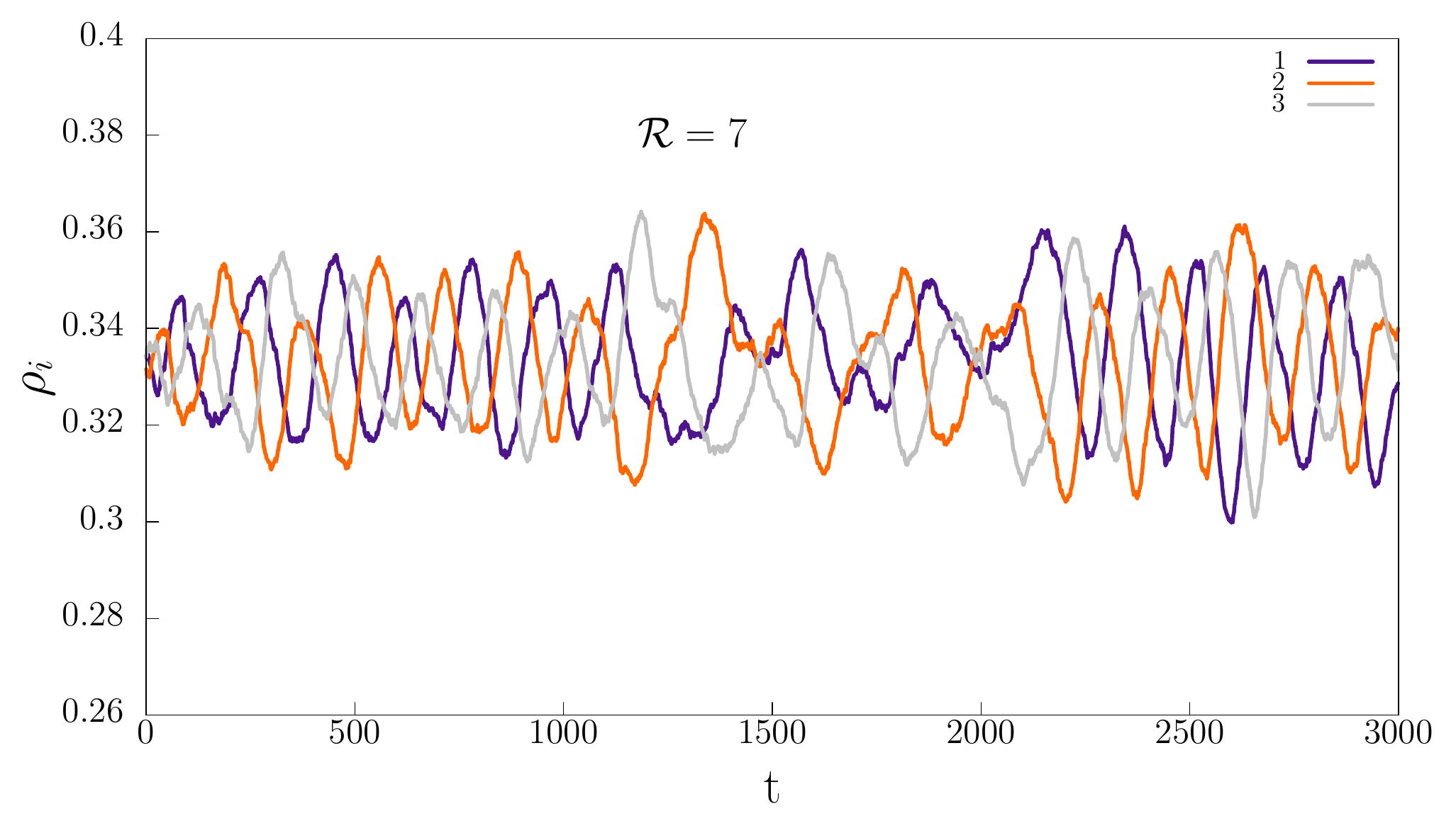}
        \caption{}\label{fig2d}
    \end{subfigure} 
\caption{Dynamics of the densities of species in simulations presented in Fig. 3. Figures a, b, and c show $\rho_i$ as a function of the time for $\mathcal{R}=0$, $\mathcal{R}=3$, and $\mathcal{R}=7$, respectively. The colours follow the scheme in Fig. 1.}
  \label{fig2}
\end{figure}
\section{Methods}
\subsection{The Stochastic Model}
We study a nonhierarchical tritrophic system whose predator-prey interactions follow the popular rock-paper-scissors game rules. 
The labeling assumed to identify the species is $i$ with $i= 1,...,3$, with the cyclic identification $i=i\,+\,3\,\beta$ where $\beta$ is an integer. According to this definition, organisms of species $i$ prey upon individuals of species $i+1$. 
In our model, organisms move directionally towards the direction with more conspecifics. 
We assume that individuals may 
scan their environment to discover the location of their conspecifics, thus moving towards them.

Our simulations are performed in a square lattice with periodic boundary conditions, following predation and mobility rules. We assumed the Lotka-Volterra implementation, which implies that the total number of individuals is conserved \cite{doi:10.1021/ja01453a010,Volterra}. Each grid point contains one individual; thus, the total number of individuals is $\mathcal{N}$, the total number of grid points. The initial conditions are built by distributing each individual at a random grid point. As time passes, interactions stochastically change the spatial configuration of individuals. 

The interactions are implemented as follows:
\begin{itemize}
\item 
Predation: $ i\ j \to i\ i\,$, with $ j = i+1$: every time one predation occurs, the grid point occupied by the individual of species $i+1$ is occupied by an offspring of species $i$.
\item 
Mobility: $ i\ \odot \to \odot\ i\,$, where $\odot$ means either an organism of any species: an individual of species $i$ switches positions with another individual of any species.
\end{itemize}
Predation and mobility interactions occur with probabilities $p$ and $m$, respectively, which are the same for all individuals of every species. Figure 1 illustrates the spatial interactions that we implemented by assuming the von Neumann neighbourhood, i.e., individuals may interact with one of their four nearest neighbours. The purple, orange, and grey arrows show that organisms of species $i$ prey upon individuals of species $i+1$; red bars indicate two organisms of any species switch positions with the same probability.

The simulation algorithm follows the steps: i) choosing a random active individual; ii) raffling one interaction to be executed; iii) in case of predation, drawing one of the four nearest neighbours as the prey to be consumed; iv) in case of mobility, the gregarious movement defines which immediate neighbour the active individual switches positions. If the interaction is executed, one timestep is counted. Otherwise, the steps are repeated. Our time unit is called generation, the necessary time to $\mathcal{N}$ timesteps to occur.
\subsection{Implementation of the Gregarious Movement}

To implement the aggregation behaviour, we first define a conditioning factor, $\alpha$, which characterising the readiness to execute the movement strategy. This quantifies the physical ability to adapt to the collective tactic or the stage of the learning process: cognitive and physiological organism's particular features. Once the individual is conditioned to move gregariously, the code proceeds the 
following the steps: i) defining a perception radius, $\mathcal{R}$, to represent the maximum distance an individual can scan the environment to be aware of the position of its conspecifics; ii) implementing a circular area for the predator to scan the vicinity (a disc of radius $\mathcal{R}$, centred in the active individual); iii) separating the observing disc into four circular sectors, each section in the directions of the one nearest neighbour (the von Neumann neighbourhood defines the immediate vicinity); 
iv) counting the number of conspecifics within each circular sector; organisms on the circular sector borders are assumed to be part of both circular sectors; v) choosing the circular sector that contains more conspecifics; in the event of a tie, making a draw between the tied directions; vi) switching positions of the active individual with the immediate neighbour in the direction of the selected circular sector. 

Figure 2 illustrates how the circular sectors are implemented for the case $\mathcal{R}=5$: the organism positioned at the black grid site switches position with the individual located in the grey background point in the direction with more conspecifics - following von Neumann's neighbourhood implementation. For selecting the direction to move, the organism scans its conspecifics in the grid sites on the North (white, light blue, and light red), south (yellow, green, and orange), east (red, pink, and orange), and west (blue, green, and light blue) directions.
To implement the perception radius $R$, our algorithm 
calculates the Euclidean distance between the active individual and the organisms in their neighbourhood
 \cite{Moura,Anti1,anti2}. This means that an active individual located at the spatial grid position ($i_A, j_A$) perceives all neighbours in the grid sites ($i_B, j_B$), with
$(i_A - i_B)^2 + (j_A - j_B)^2 \leq R^2$.
\subsection{Spatial Autocorrelation Function}

To study how organisms of the same species are spatially correlated, we calculate the autocorrelation function from the inverse Fourier transform of the spectral density as
\begin{equation}
C(\vec{r}') = \frac{\mathcal{F}^{-1}\{S(\vec{k})\}}{C(0)},
\end{equation}
where $S(\vec{k})$ is given by
\begin{equation}
S(\vec{k}) = \sum_{k_x, k_y}\,\varphi(\vec{\kappa}),
\end{equation}
and $\varphi(\vec{\kappa})$ is Fourier transform
\begin{equation}
\varphi(\vec{\kappa}) = \mathcal{F}\,\{\phi(\vec{r})-\langle\phi\rangle\}.
\end{equation}

The function $\phi(\vec{r})$ represents the spatial distribution of individuals of species $1$ ($\phi(\vec{r})=0$ and $\phi(\vec{r})=1$ indicate the absence and the presence of an individual of species $1$ in at the position $ \vec{r}$, respectively). Therefore, the spatial autocorrelation function is given by
\begin{equation}
C(r') = \sum_{|\vec{r}'|=x+y} \frac{C(\vec{r}')}{min (2N-(x+y+1), (x+y+1)}.
\end{equation}
We then use the autocorrelation function to find the scale of the spatial domains as $C(l)=0.15$, where $l$ is the characteristic length.
\subsection{Densities of species}
To compute how the emergence of single-species spatial domains due to the aggregation strategy affects the population dynamics, we
calculate the densities of species $\rho$, i.e., the fraction of the grid occupied by individuals of the species $i$, that is a function of time $t$, i.e., $\rho_{i}(t) = I_1(t)/\mathcal{N}$.
\begin{figure}
\centering
    \begin{subfigure}{.2\textwidth}
        \centering
        \includegraphics[width=36mm]{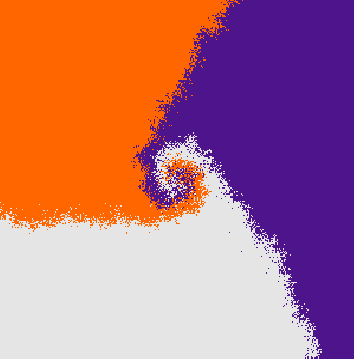}
        \caption{}\label{fig4b}
    \end{subfigure} %
           \begin{subfigure}{.2\textwidth}
        \centering  
        \includegraphics[width=36mm]{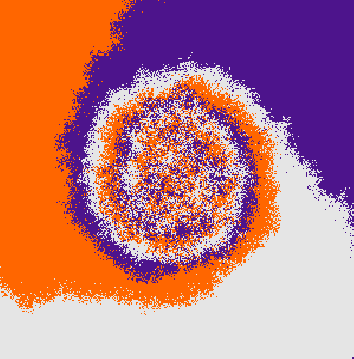}
        \caption{}\label{fig4c}
    \end{subfigure}\\
    \begin{subfigure}{.2\textwidth}
        \centering
        \includegraphics[width=36mm]{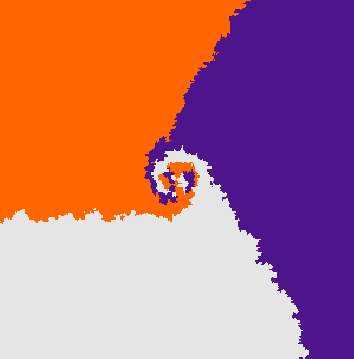}
        \caption{}\label{fig4e}
    \end{subfigure}
       \begin{subfigure}{.2\textwidth}
        \centering
        \includegraphics[width=36mm]{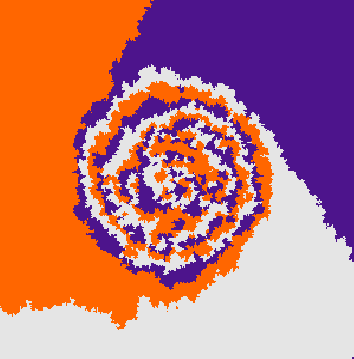}
        \caption{}\label{fig4f}
    \end{subfigure} %
\caption{Snapshots captured from the simulation of a circular spiral growth in a lattice with $300^2$ grid points without periodic boundary conditions.
Figures a, and b show the spatial configuration for the standard model at $t=55$ and $t=148$ generations.Figures c and d depict the circular spiral at $t=170$ and $t=520$ generations, for the modified model where organisms aggregate with a perception radius $R=5$.}
  \label{fig2}
\end{figure}
\begin{figure}[t]
\centering
\includegraphics[width=87mm]{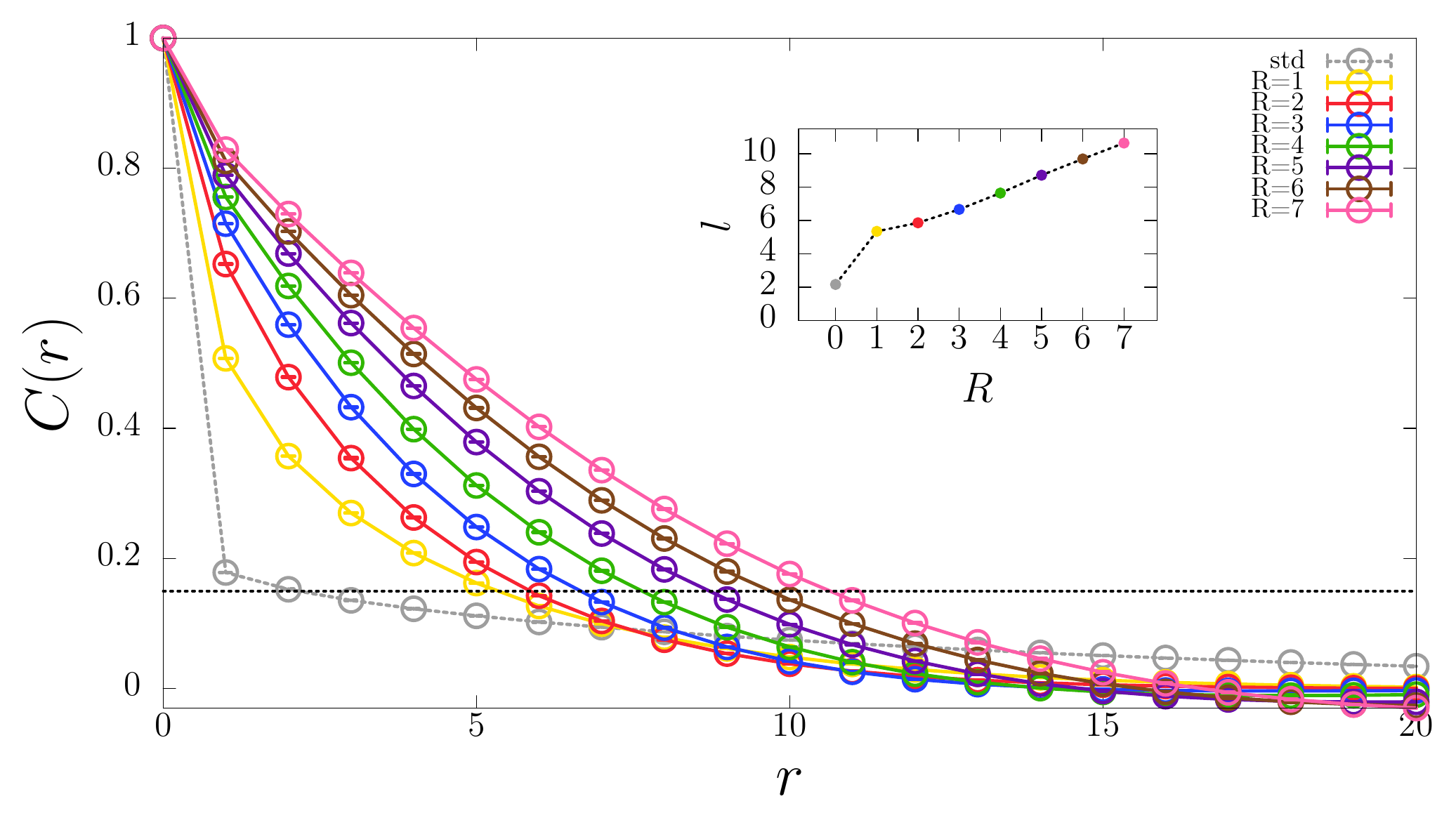}
\caption{Autocorrelation functions $C$ in terms of the perception radius. Grey indicate the outcomes for the standard model, while yellow, red, blue, green, purple, brown, and pink show the results for $\mathcal{R}=1$ to $\mathcal{R}=7$, respectively. 
The error bars indicate the standard deviation. The horizontal black line indicates the threshold assumed to calculate the characteristic length, depicted in the inset as a function of 
$\mathcal{R}$.}
	\label{fig3}
\end{figure}
\subsection{Predation Risk}

Finally, we explore the impact of the aggregation strategy on the risk of an individual being consumed by predators in one generation interval.
For this purpose, the algorithm follows the steps: i) counting the total number of individuals of species $i$ at the beginning of each generation; ii) computing the number of individuals of species $i$ are preyed on during the generation; iii) calculating the predation risk, $\zeta_i$, with $i=1,2,3$, as the ratio between the number of consumed individuals and the initial amount.


\section{Pattern Formation}

We first investigated a scenario where organisms of every species are thoroughly conditioned to perform the collective aggregation strategy. 
The first step was to observe the effects of the grouping behaviour on the organisms' spatial organisation. For this purpose, we first run a single realisation of the standard model in a lattice with $300^2$ for a timespan of $3000$ generations - in this case, all organisms move randomly.
Subsequently, we performed simulations where all organisms of every species move gregariously, considering the perception radii $\mathcal{R}\,=\,3$ and $\mathcal{R}\,=\,7$. All simulations were performed with $p\,=\,0.25$ and $m\,=\,0.75$. 

Figure $3$ shows that aggregation strategy leads to pattern formation, with organisms of the same species occupying separated spatial domains. The colours show individuals according to the scheme in Fig. $1$. The cyclic dominance in the predator-prey interactions generates waves, where predators invade territories dominated by prey. According to the snapshot in Fig. 3a, only the cyclic dominance of the predator-prey interactions described by the rock-papers-scissors rule is insufficient for the emergence of departed single-species domains. This happens only if the antipredator behaviour leads organisms to move gregariously. Furthermore, perceiving further, an individual can accurately identify the direction with a larger group of conspecifics; otherwise, the probability of inadvertently joining a smaller group closer to it is high. Because of this, the average size of single-species domains increases with $R$, as one sees in the snapshots Figs. 3b and 3c.

Figure 4 shows the dynamics of the densities of species for the simulations in Fig. 3. Figure 4a shows that the spatial dominance in the standard model (random movement) is cyclic because of the predator-prey rules. This happens despite the irregular pattern formation shown in Fig. 3a. However, the outcomes reveal that the frequency of the densities of species decreases if organisms move gregariously, forming clumps. According Figs. 4b and 4c show that the benefits of the aggregation strategy accentuate if the organism can perceive further, thus creating bigger groups. In this case, predators can access only prey on the border of single-species domains, thus decreasing the predation activity and, consequently, the frequency of spatial densities.

To observe the pattern formation more closely, we ran one simulation starting from a particular spatial configuration with three single-species domains symmetrically disposed on the lattice - the angles at the vertex are initially $2\pi/3$.
For this simulation, the grid periodic boundary conditions were relaxed.
Figures 5a and 5b show snapshots of the lattice for the standard model, where organisms move randomly for $t=55$ and $t=148$ generations, respectively. Figures 5b and 5d depict the organism' spatial displacement in the case of aggregation strategy after $t=170$ and $t=520$ generations, respectively. Both simulations were performed in square lattices with $300^2$ grid sites, for $p=0.25$ and $m=0.75$; in the case of the aggregation antipredator tactic, it was assumed $\mathcal{R}\,=\,5$. The purple, orange, and grey regions are occupied by species $1$, $2$, and $3$, respectively. 

As soon as the simulation starts, the area where all species are in contact starts spiralling, creating a circular spiral that grows until occupying the entire grid. This is caused by the cyclic predator-prey interactions
described in Fig. $1$. Our outcomes show that the local dynamics of the rock-paper-scissors rules without the aggregation antipredator tactics allow all species to mix themselves and form irregular groups in the circular spiral core. In contrast, the gregarious movement creates departed spatial domains inside the circular spiral, keeping the pattern after the circular spiral occupying the entire grid. One consequence of the internal pattern formation caused by the aggregation is a delay in the circular spiral growth: it took more than three times to fill the entire grid when the aggregation tactic was executed. The delay in the local predator-prey dynamics is more significant for larger organisms' perception radii.

\begin{figure}
\centering
\includegraphics[width=77mm]{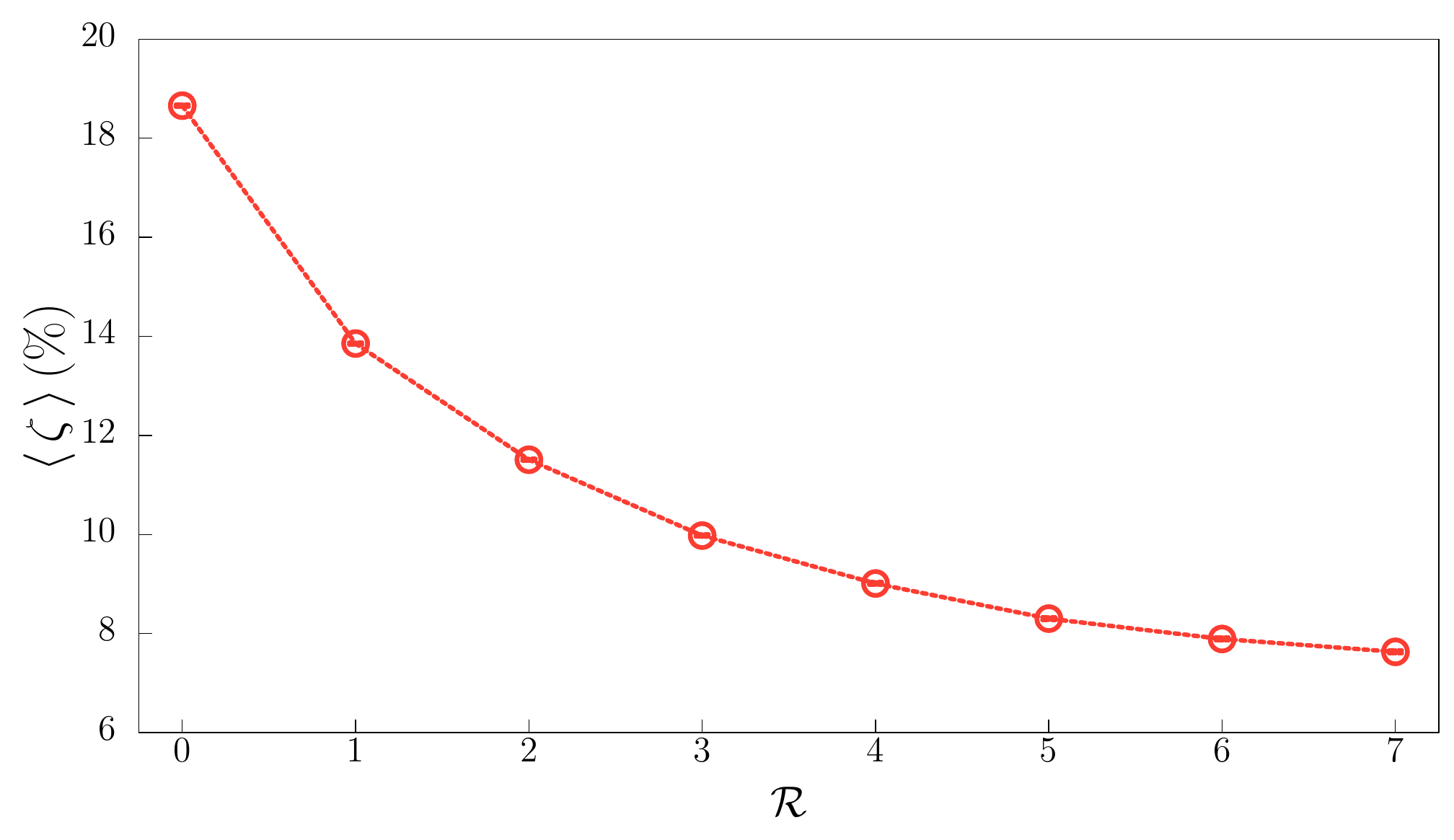}
\caption{Predation risk is due as a function of the perception radius in the case of all individuals of every species fully conditioned.
The red line depicts the mean value of $\zeta_i$ for sets of $100$ simulations for $\mathcal{R}\,\leq\,7$, with $\mathcal{R}=0$ representing the standard models. The error bars show the standard deviation. }
	\label{fig3}
\end{figure}
\section{Spatial Autocorrelation Function}

Now we investigate the scale of departed spatial domains formed because of aggregation strategy. Because organisms of every species are totally conditioned to perform the behavioural tactic, the average size of single-species areas is the same irrespective of the species - the interaction probabilities $p$ and $m$ are the same for all species. We then concentrate on computing the average characteristic length for domains occupied by species $1$.
Figure 6 depicts the spatial autocorrelation function in terms of the radial coordinate $r$. The outcomes were averaged from a set of $100$ simulations with different random initial conditions, running in lattices with $\mathcal{N}=300^2$. The spatial statistics were performed using the spatial configuration captured after $3000$ generations, for $p=0.25$ and $m=0.75$. The horizontal dashed black line represents the threshold considered to calculate the length scale, $C(l)\, =\, 0.15$. The standard deviation is smaller than the symbol used to depict the mean autocorrelation value.

First, we compute the autocorrelation function for the standard model, where all organisms move randomly (grey dashed line). Then, we studied the spatial agglomeration size for the antipredator strategy considering $\mathcal{R}\,\leq7$. The results for the aggregation tactic are depicted by the solid yellow, red, blue, green, purple, brown, and pink lines for $\mathcal{R}=1$, $\mathcal{R}=2$, $\mathcal{R}=3$, $\mathcal{R}=4$, $\mathcal{R}=5$, $\mathcal{R}=6$, and $\mathcal{R}=7$, respectively. The inset figure shows the characteristic length for each case, with the circle colour indicating the perception radius $\mathcal{R}$ - the standard case is represented by $\mathcal{R}=0$
Our findings reveal that the average size of spatial clumps formed by the gregarious movement grows with $\mathcal{R}$. 

\section{Predation Risk and Dynamics of the Densities of Species}

The population dynamics is defined by a cyclic territorial dominance of species $i$ ($i=1,2,3$), characteristic to the rock-paper-scissors model \cite{doi:10.1021/ja01453a010,Volterra}. The outcomes show that the frequency of cyclic dominance is lowered if organisms move gregariously, with this effect being more substantial for a larger perception radius. This is a consequence of a reduced predation activity because of the
protection that organisms benefit in larger prey groups. 

To quantify the role of perception radius on the predation risk, we performed $100$ simulations with different random initial conditions for each value of $\mathcal{R}$. To avoid the noise inherent in the pattern formation stage, we computed the predation risk considering only the second half of the simulation.
In this case, all individuals are conditioned and we assumed the same predation and mobility probabilities for every species; thus, we focus only on finding the predation risk for species $1$ because $\zeta = \zeta_i$, with $i=1,2,3$.

Figure 7 shows the mean value percentage value for the predation risk as a function of the perception radius for $0\,\leq\,\mathcal{R}\,\leq\,7$,
with $\mathcal{R}\,=\,0$ representing the standard model. The error bars indicate the standard deviation (the error bars are smaller than the symbol used to depict the mean predation risk).
The simulations were performed in lattices with $300^2$ grid points for a timespan of $3000$ generations; it was assumed $p\,=\,m\,=\,0.5$. 

Our results show that performing aggregation is an advantageous antipredator strategy in the spatial rock-paper-scissors model because:
i) individuals on the boundaries of the single-species domains do not move away from the group they belong; thus, reducing the exposure to the predator; ii) organisms within the single-species areas are topologically protected because they are out of reach of the predator. Therefore, aggregation is more profitable if individuals can scan further distances, thus creating larger conspecific groups (according to the inset of Fig. 6).

\section{Role of the Conditioning Process}
We investigate the general case where not all organisms are conditioned to perform the aggregation strategy considering two scenarios:
i) species $1$ has more organisms conditioned to move gregariously than the other species: $\alpha_1\,\geq\,\alpha_2\,=\,\alpha_3$; ii) species $1$ has less individuals conditioned than the other species: $\alpha_1\,\leq\,\alpha_2\,=\,\alpha_3$ (where $\alpha_i$ represents the conditioning factor of species $i$, with $i=1,2,3$).

To observe the impact of the unevenness in the organisms' ability to form clumps in the spatial patterns, we run single simulations in grids with $300^2$ sites, for $\mathcal{R}=7$, $p=0.25$, and $m=0.75$. Figure 8a shows the result for the case where all individuals of species $1$ are conditioned while organisms of species $2$ and $3$ cannot move gregariously ($\alpha_1 =1$ and $\alpha_2 =\alpha_3=0$).
The large number of purple clumps in Fig. 8a reveal that aggregation is a good self-protection strategy in cyclic models, resulting in territorial dominance of species $1$ in detriment of the population decline of species $3$. Running $100$ simulations with different initial conditions using the same lattice size and parameters of the single simulation in Fig. 8a, we found that the characteristic lengths for the single-species domains: $l_1\,=\,6.59\pm0.034$, $l_2\,=\,6.92\pm0.032$, and $l_3\,=\,3.40\pm0.00072$. 

The opposite case is shown in Fig. 8b, where the individuals of species $1$ are the only ones not conditioned ($\alpha_1 =0$ and $\alpha_2 =\alpha_3=1$). In this case, species $1$ is at a disadvantage in the cyclic game: species $3$ benefits from the random mobility of species $1$ to proliferate and create large areas (grey regions), making it difficult to be caught by organisms of species $2$. We quantified the scale of group species in Fig. 8b by averaging the results from a set of $100$ simulations with different initial conditions revealed that $l_1\,=\,9.81\pm0.057$, $l_2\,=\,12.658\pm0.166$, and $l_3\,=\,14.22\pm0.083$.

To observe how predation risk and densities of species depend on the level of conditioning of organisms of species $1$, we performed sets of $100$ simulations for $0\,\leq\,\alpha_1\,\leq\,1$, in intervals of $\delta\,\alpha\,=\,0.05$. First, the purple lines in Figs. 9a and 9b indicate that the more the fraction of conditioned organisms, the more profitable is the aggregation as an antipredator strategy for species $1$: the effect of the reduction in the predation risk resulting in population growth accentuates as more organisms learn the strategy. The outcomes also show that the grouping of organisms of species $1$ also benefits species $2$ since the more individuals of species $1$ is aggregating, the fewer individuals are moving towards regions with a high concentration of species $2$. However, the reduction of predation risk of species $2$ observed in Fig. 9a is not reflected in a high spatial density of species $2$ because of the low density of species $3$. 

Figure 10a shows the variation of the predation risk in terms of $\alpha_1$, for all organisms of species $2$ and $3$ are conditioned. As fewer organisms of species $1$ are ready to move gregariously, predation risk increases, resulting in a small population. In this scenario, species $3$ profits more because organisms of species $1$ are not in groups. According to Figure 10b, the consequence is the reduction of predation risk of species $3$ because most of the individuals of species $3$ are topologically protected inside spatial domains formed when individuals aggregate. Another consequence is that as the population of species $3$ grows, more individuals of species $2$ are consumed, causing the increase of predation risk of species $2$.

\begin{figure}
    \centering
    \begin{subfigure}{.2\textwidth}
        \centering
        \includegraphics[width=36mm]{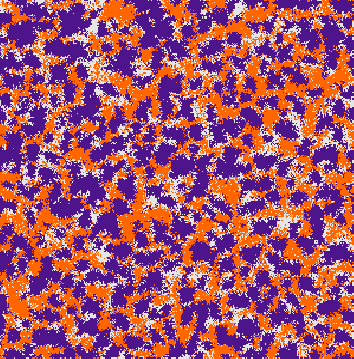}
        \caption{}\label{fig4b}
    \end{subfigure} %
           \begin{subfigure}{.2\textwidth}
        \centering  
        \includegraphics[width=36mm]{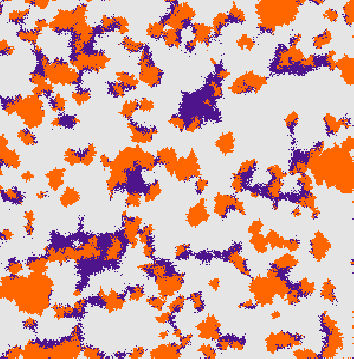}
        \caption{}\label{fig4c}
    \end{subfigure}
\caption{Snapshots from simulations for uneven conditioning for the aggregation strategy. The realisations ran in lattices in $300^2$ sites for a timespan of $3000$ generations, assuming $\mathcal{R}=7$.
Figures a shows the spatial patterns for the case where only organisms of species $1$ move gregariously, while Fig. b depicts the spatial organisation where only organisms of species $1$ are not conditioned to aggregate. The error bars show the standard deviation; the colours follow the scheme in Fig. 1.}
  \label{fig2}
\end{figure}

\begin{figure}
 \centering
       \begin{subfigure}{.4\textwidth}
        \centering
        \includegraphics[width=75mm]{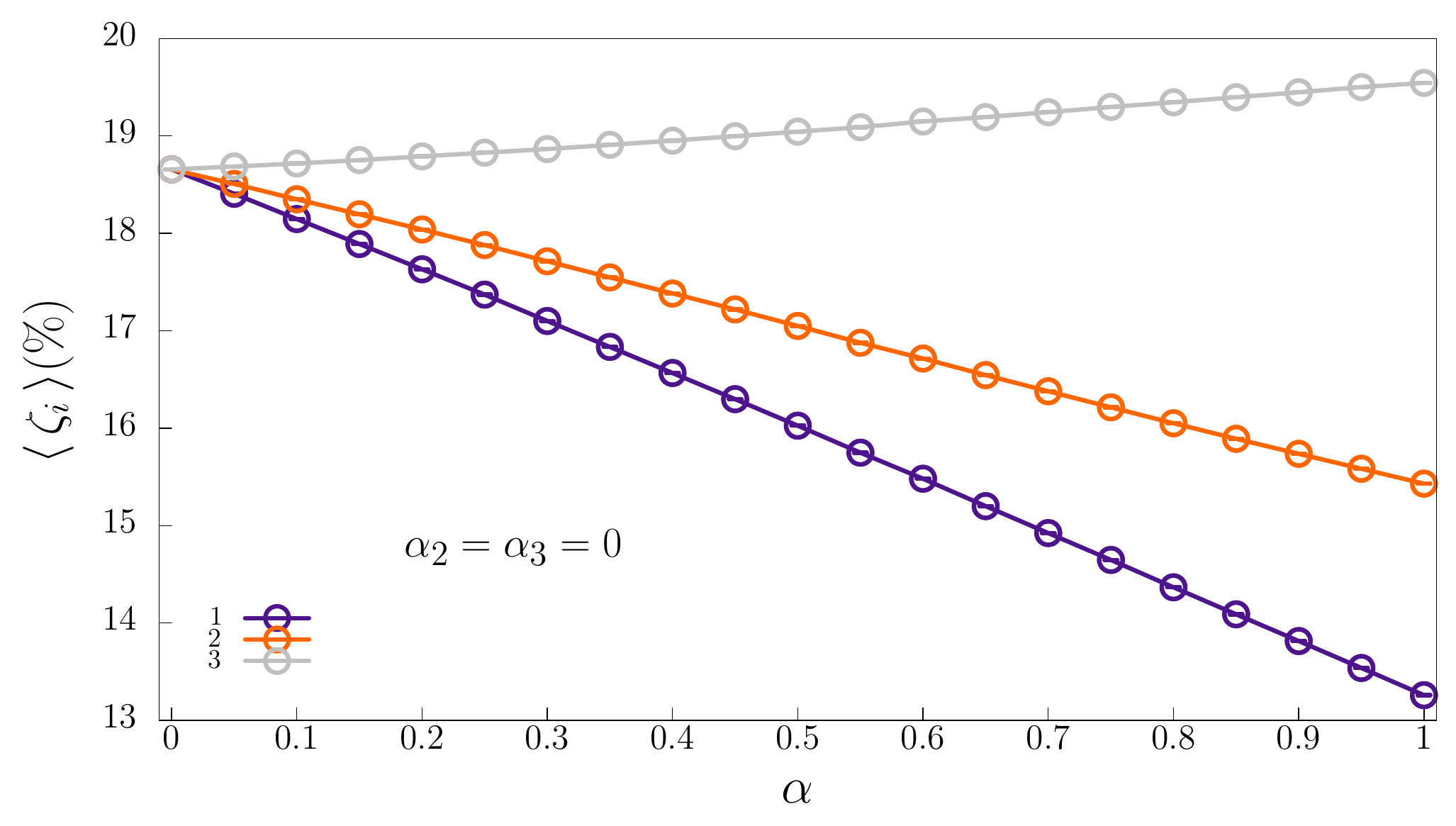}
        \caption{}\label{fig2b}
    \end{subfigure}\\
   \begin{subfigure}{.4\textwidth}
        \centering
        \includegraphics[width=75mm]{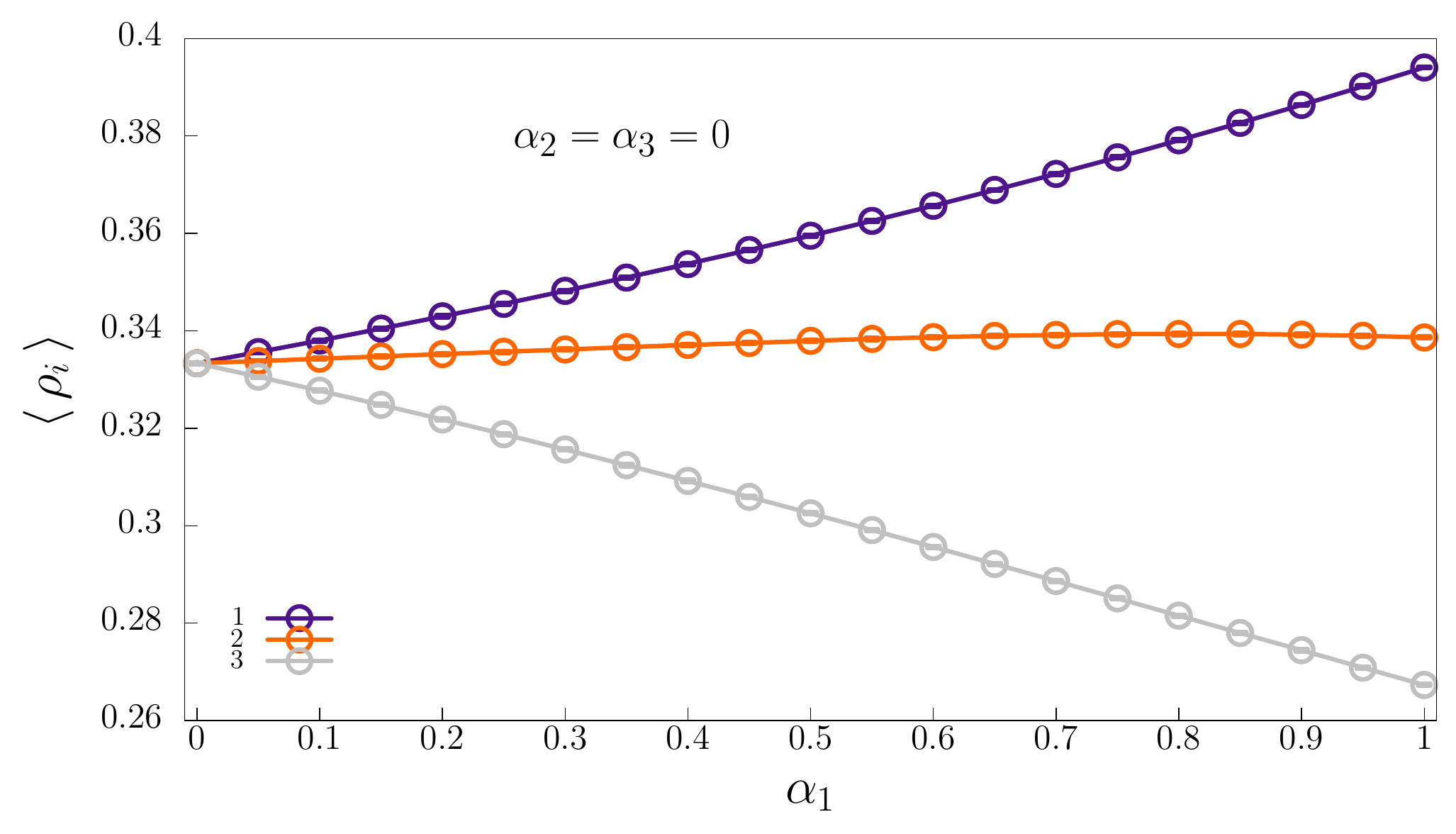}
        \caption{}\label{fig2d}
    \end{subfigure} 
\caption{Predation risk (Fig. a) and densities of species (Fig. b) as a function of $\alpha_1$, for $\alpha_2\,=\,\alpha_3\,=0$.
The outcomes were obtained from sets of $100$ simulations for each value of $\alpha_1$, in lattices with $300^2$ grid points for $\mathcal{R}\,=\,3$. The error bars show the standard deviation; the colours follow the scheme in Fig. 1.}
  \label{fig2}
\end{figure}
\begin{figure}
 \centering
       \begin{subfigure}{.4\textwidth}
        \centering
        \includegraphics[width=75mm]{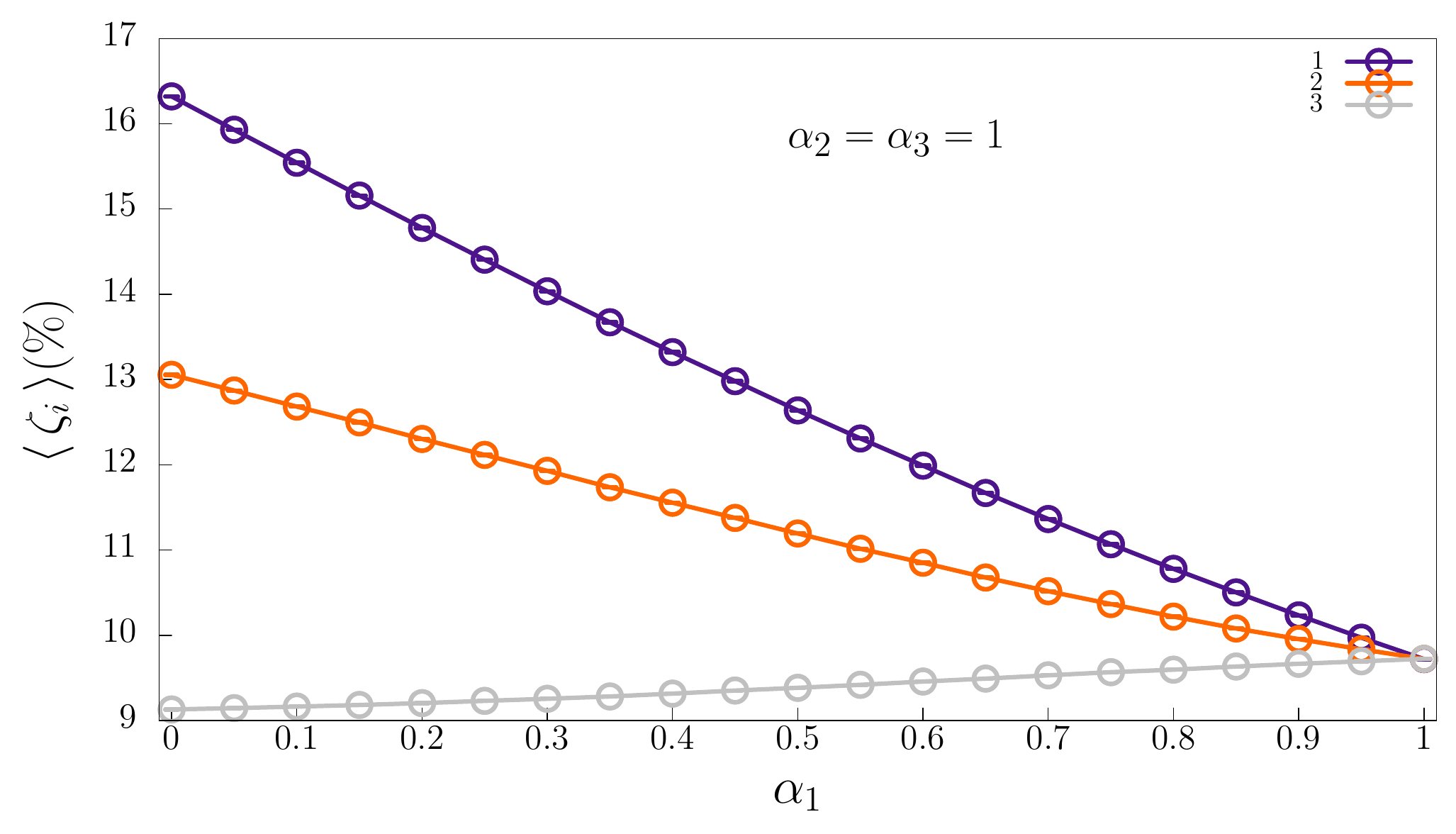}
        \caption{}\label{fig2b}
    \end{subfigure}\\
   \begin{subfigure}{.4\textwidth}
        \centering
        \includegraphics[width=75mm]{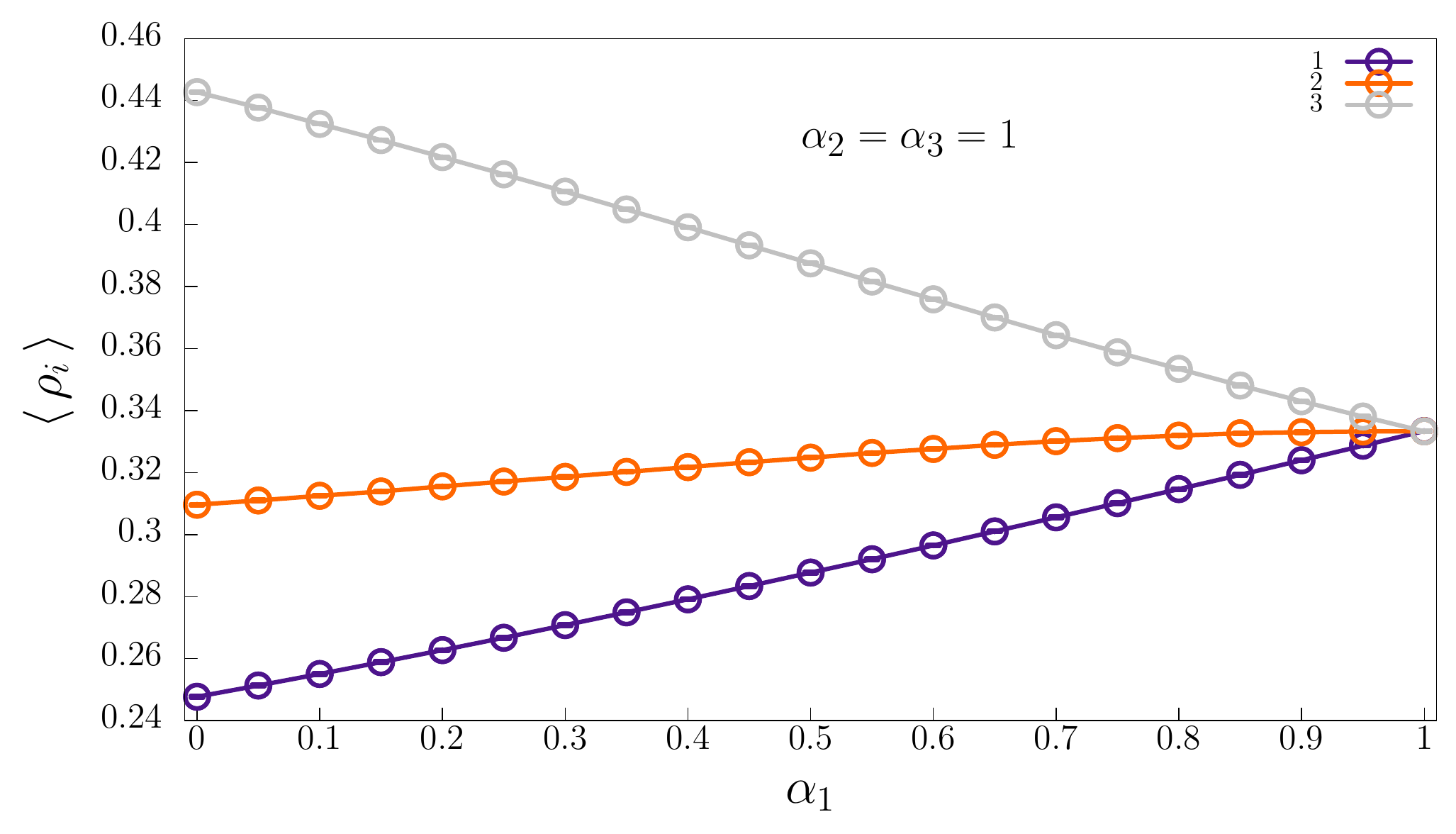}
        \caption{}\label{fig2d}
    \end{subfigure} 
\caption{Predation risk (Fig. a) and densities of species (Fig. b) in terms of $\alpha_1$,  for $\alpha_2\,=\,\alpha_3\,=\,1$.
The results were averaged from sets of $100$ simulations in lattices with $300^2$ grid points for $\mathcal{R}\,=\,3$. The error bars show the standard deviation; the colours follow the scheme in Fig. 1.}
  \label{fig2}
\end{figure}


\section{Conclusions}
\label{sec6}

We study a cyclic game where the rock-paper-scissors game rules describe predator-prey interactions. As an antipredator strategy, organisms may form groups to minimise the chances of being caught by a nearby predator. The behavioural movement of aggregation can be performed correctly only if an individual has the physical and cognitive abilities to distinguish its neighbours, identifying the direction where the larger number of conspecifics are. 
Performing stochastic agent-based simulations, we found that the gregarious movement leads to the emergence of single-species domains that are not present in the standard model. This gives the individuals an efficient refuge because predators can only catch prey on the borders of the agglomerations - the larger the characteristic length of the single-species domains, the lower is the predation risk. Our results show that the topological advantage increases if organisms can perceive further the neighbourhood, thus choosing more accurately the best direction to move.

Our findings show that, in general, the more conditioned organisms, the more dominant is the species in the cyclic spatial game. The consequence is a higher spatial species for the more conditioned species. In opposition, if one species is less conditioned, its predation risk increases, allowing its predator to control the territory. Our statistics results show that the more significant is the unevenness in the number of conditioned organisms, the more accentuated the disequilibrium in territorial control.

Our results can be extended to generalise the rock-paper-scissors model an arbitrary number of species $N$, where organisms of species $i$, with $i=1,2,3,...,N$ prey upon individuals of species $i+1$. If organisms of every species can move gregariously, the emerging single-species domains influence the dynamics of the spatial patterns differently for $N\,\geq\,4$. For $N=3$, organisms of species $i$ perform predator-prey interactions with organisms of all other species, being predators of individuals of species $i+1$ and prey for species $i-1$. This is not valid for 
$N=4$ because individuals of species $i$ are not predators nor prey from organisms of species $i+2$. The consequence is that extra protection is provided to the organisms of species $i$ that form a group within an agglomeration of individuals of species $i+2$. In general, the number of the non-interacting single-species domains increases for larger $N$ - no predator-prey interaction between individuals of species $i$ and $i+\kappa$, with $i-2\,\leq\,\kappa\,\leq\,1+2$.  

Our implementation of the gregarious movement 
allowed us to conclude how aggregation works to reduce the organisms' predation risk.  Our algorithm implements what happens, for example, in mites species,
where each individual learns to perceive the odour of their conspecifics \cite{Chemical0,Chemical1}. Once detected the origin of the more intense chemical signal is received, the organism moves towards its direction.  However, it is possible to 
introduce new variables to model an adaptive behaviour where organisms interpret the neighbourhood's signals more
accurately.  For example, by analysing the chemical signals that indicate the presence of prey and predators in the neighbourhood, an individual can decide if aggregation is the best movement strategy at the moment.

Overall, our findings show that the aggregation strategy brings positive results for species in spatial cyclic models. Our discoveries may also be helpful to the biologists to comprehend systems where adaptive processes are responsible for biodiversity stability.

 
\section*{Acknowledgments}
We thank CNPq, ECT, Fapern, and IBED for financial and technical support.

\bibliographystyle{elsarticle-num}
\bibliography{ref}

\end{document}